\documentclass[conference]{IEEEtran}
\usepackage[pdftex]{graphicx}

\ifCLASSINFOpdf
\else
\fi
\usepackage[noend]{algpseudocode}
\usepackage{algorithm, algorithmicx}
\usepackage{layouts}
\usepackage{subfig}
\usepackage{amsmath,amssymb,amsfonts}

\usepackage[font=small]{caption}
\usepackage{graphicx}
\usepackage{textcomp}
\usepackage{mathtools}
\usepackage{mathtools}

\usepackage[bookmarks=false]{hyperref}

\usepackage{xcolor}

\begin{document}
\title{
Optimizing Layer-Fused Scheduling of Transformer Networks on Multi-accelerator Platforms}

\author{\IEEEauthorblockN{Steven Colleman*, Arne Symons*, Victor J.B. Jung$\dagger$, Marian Verhelst*}
\IEEEauthorblockA{*Department of Electrical Engineering, MICAS-ESAT, KU Leuven, Belgium.}
\IEEEauthorblockA{$\dagger$Integrated Systems Laboratory, ETH Zürich, Switzerland.}
}

\maketitle
\begin{abstract}
The impact of transformer networks is booming, yet, they come with significant computational complexity. It is therefore essential to understand how to optimally map and execute these networks on modern neural processor hardware. So far, literature on transformer scheduling optimization has been focusing on deployment on GPU and specific ASICs. This work enables extensive hardware/mapping exploration by extending the DSE framework Stream towards support for transformers across a wide variety of hardware architectures and different execution schedules. After validation, we explore the optimal schedule for transformer layers/attention heads and investigate whether layer fusion is beneficial to improve latency, energy or memory requirements. Our study shows that the memory requirements for active feature data can be drastically reduced, by adapting the execution schedule based on the size of the input of the attention head. 
\end{abstract}

\section{Introduction}
\IEEEPARstart{N}{owadays}, machine learning algorithms are gaining more and more importance in a broad range of daily-life applications in fields like natural language and image processing. Over the past years, transformer networks have been introduced \cite{vaswani2017attention}, with a fast growing set of applications \cite{transformers}. These networks use the mechanism of self-attention, giving a data-dependent weight to each element of the network input data. 

\begin{figure}[btp]
\centering
\includegraphics[width=1.0\linewidth]{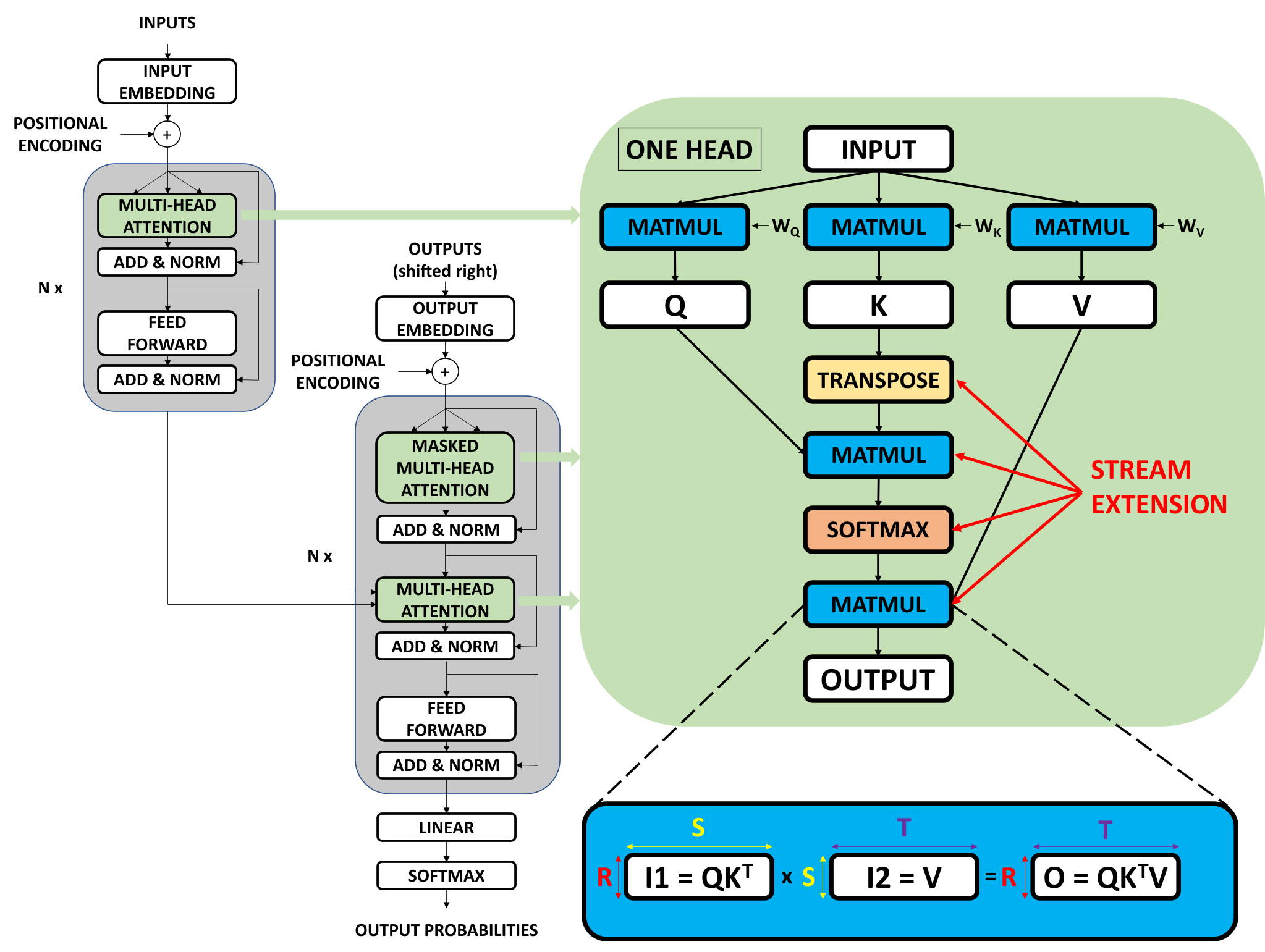}
\caption{Operators of a transformer network, highlighting one attention head containing 5 matrix-matrix multiplications (3x features$\times$weights, 2x features$\times$features), a transpose and a softmax layer. Stream \cite{symons2022towards} is extended with these additional layer types. 
}
\label{fig:shape}
\vspace*{-0.3cm}
\end{figure}

These algorithms, however, lead to new challenges in terms of efficient hardware mapping, as transformer networks contain layer types and sequences not present in traditional neural networks. Many practical implementations of transformer-optimized processors have been proposed 
\cite{wang202228nm}\cite{tu202228nm}\cite{i̇slamoğlu2023ita}. Yet, they are ad hoc designs, often focusing on a specific network. Alternative studies explore the optimal execution schedule, assessing the potential benefits of layer fusion of specific transformer layers 
\cite{kao2023flat}\cite{sheng2023high}, yet do so for specific hardware architectures and networks. 
The work of \cite{colleman2022optimizing} performs a broader analysis, but limits the transformer scheduling space to layer-by-layer execution only. The goal of this paper is, therefore, to approach the design and scheduling exploration more methodologically, by creating an analytical model to optimize the mapping efficiency of transformer networks on a wide variety of neural processing hardware architectures. The framework should not only support layer-by-layer schedules, but also the emerging scheduling technique of deep layer-fused execution \cite{symons2022towards}. 

For traditional CNNs, execution efficiency has been analyzed and optimized by making use of analytical frameworks such as Timeloop \cite{parashar2019timeloop}, ZigZag \cite{mei2021zigzag}, Kwon et al. \cite{kwon2021heterogeneous}. DeFiNES \cite{mei2022defines} extended the scheduling space to deep layer-fused execution for single-core architectures and Stream \cite{symons2022towards} extended this further towards multi-core execution. However, none of these frameworks support transformer networks, lacking support for transpose and softmax layers. Therefore, this work fills this gap by creating an analytical cost estimation framework which can both handle transformer networks as well as multi-core layer-fused execution.

Section~\ref{methodology} first sketches the background on transformers and Stream's analytical cost estimation, followed by introducing the core methodology of this paper: the requirements to accurately model layer-fused execution of transformers.

Section~\ref{validation} contains a validation of the extended framework against real-world hardware to verify the correct functionality of the modified framework. 

Finally, Section~\ref{exploration} uses the developed framework to perform explorations on efficiently mapping transformer networks on various hardware architectures, in order to explore the optimal schedule for transformer layers/attention heads and investigate whether layer fusion is beneficial to improve latency, energy or memory requirements.

\section{Background \& Methodology}
\label{methodology}

\subsection{Background: Transformer Networks}
Encoder-decoder transformer networks contain several attention heads~\cite{vaswani2017attention}, illustrated in Fig.~\ref{fig:shape}, in which scale dot-product attention takes place:

An input matrix of size $M \times N$ is multiplied with three different weight matrices $W_Q$, $W_K$ and $W_V$ of size $N \times N$ to obtain three new matrices $Q$ (query), $K$ (key) and $V$ (value), respectively, all of size $M \times N$. Based on these three matrices, the attention output is calculated as:

\begin{equation}
    Attention(Q, K, V) = softmax \left( \frac{QK^T}{\sqrt{d_k}} \right) V
\end{equation}
Here, $d_k$ is a constant and will therefore not be taken into account in this work as in practice, this division can be done by adjusting the weights from one of the three mentioned weight matrices. This means that the attention block can be described with 7 layers: 5 matrix-matrix multiplications, a transpose layer, and a softmax layer. This softmax operation happens row-wise. With input matrix $QK^T$ of size $M \times M$, the output of the softmax layer will be defined as:
\begin{equation}
softmax(QK^T_{(i,j)}) = \frac{exp(QK^T_{(i,j)})}{\sum_{j = 1..M} exp(QK^T_{(i,j)})}
\end{equation}
In practice, transformer networks make use of multi-headed self-attention, implying that every attention layer consists of multiple previously-described heads in parallel. 

Throughout this text, the left and right input matrices of a matrix-matrix multiplication will be denoted by $I1$ and $I2$, respectively, with $O$ being the output matrix, with following dimensions: $I1(R \times S), I2(S \times T), O(R \times T)$. Therefore, for the computation of $Q$, $K$ and $V$, $R = M$ and $S = T = N$. For the computation of $QK^T$, $R = T = M$ and $S = N$. For the computation of $QK^T.V$, $R = S = M$ and $T = N$.

\subsection{Background: Stream} 
The open-source framework Stream~\cite{symons2022towards} is a hardware estimation framework that evaluates the (energy, latency, memory requirement) performance of executing a given neural network, with a specific execution schedule on a specific hardware architecture. 

\begin{figure}[btp]
\centering
\includegraphics[width=1.0\linewidth]{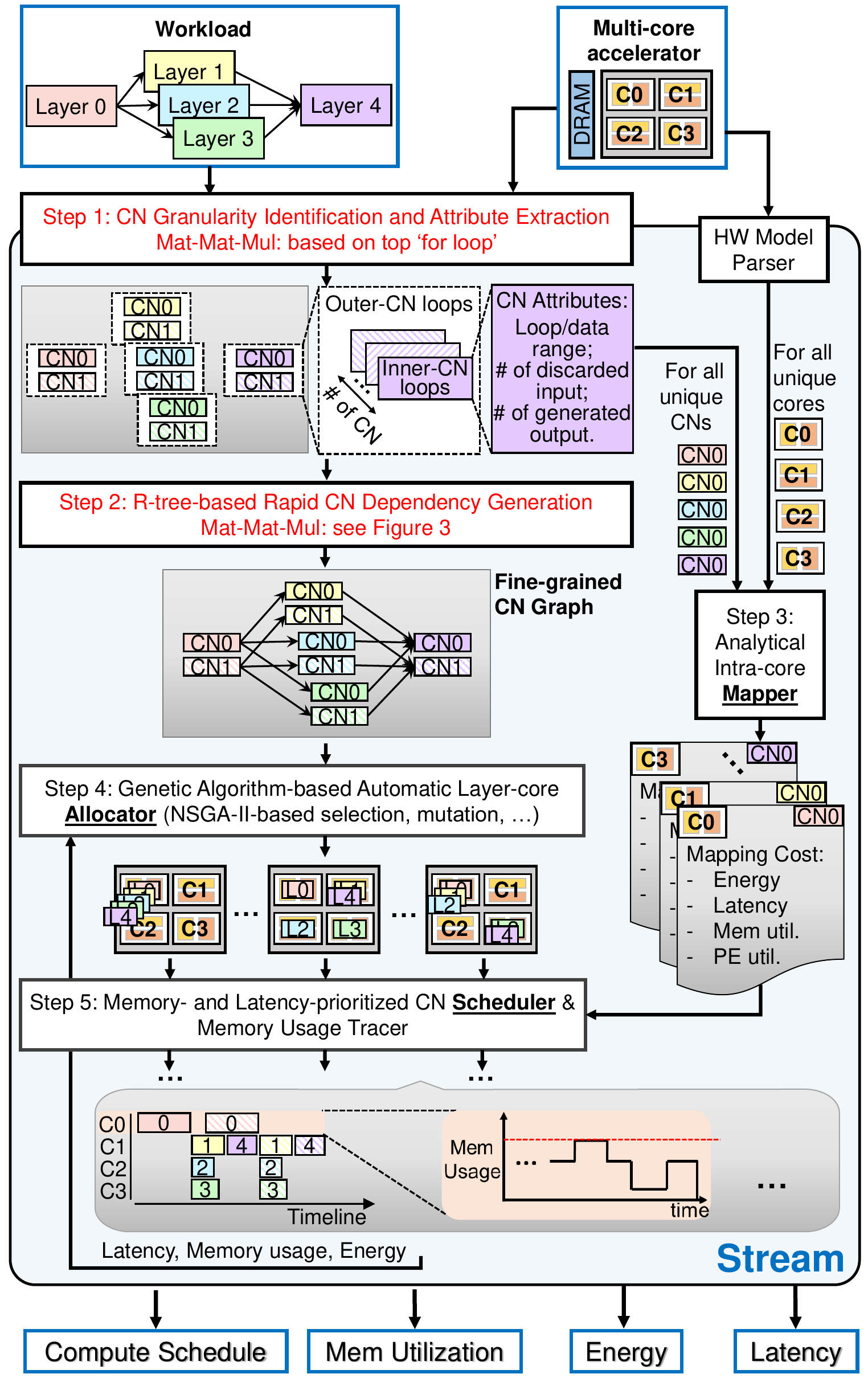}
\caption{Illustration of Stream \cite{symons2022towards}, with indicated in red the blocks where adaptations have been made.}
\label{fig:streamoriginal}
\vspace*{-0.3cm}
\end{figure}

Stream works in 5 steps, illustrated in Fig. \ref{fig:streamoriginal}:
\begin{enumerate}
\item The framework splits the layers of the network into fine-grained computation nodes. Without splitting, only layer-by-layer execution would be possible. These form the basis for the layer-fused scheduling of step 5.
\item An R-tree-based dependency generator derives for each computation node on which computation node(s)' output of the previous layer(s) it depends. These dependencies are needed to explore layer-fused scheduling~\cite{goetschalckx2019breaking}, which exploits the fact that the partial output data of a given layer can already be used in the execution of a following layer before the execution of the former layer is completely finished. 
\item An analytical intra-core mapping where the mapping of each computation node on each core of the hardware architecture is optimized for latency and/or energy. The spatial unrolling and temporal mapping of the node are optimized~\cite{symons2021loma}. With spatial unrolling, we mean the physical parallelization over the PE array for improved utilization within a clock cycle. With temporal mapping, we mean the order and splitting of the for loops for better reuse across clock cycles. 
\item A genetic algorithm optimizes which layer should be allocated to which core.
\item A scheduling algorithm decides in which order the computation nodes should be executed to optimize overall latency or memory usage with a configurable cost function.
The framework iterates between steps 4 and 5 to find the optimal allocation and scheduling.
\end{enumerate}

\subsection{Proposed Transformer Extensions for Stream}
\begin{figure}[btp]
\centering
\includegraphics[width=1.0\linewidth]{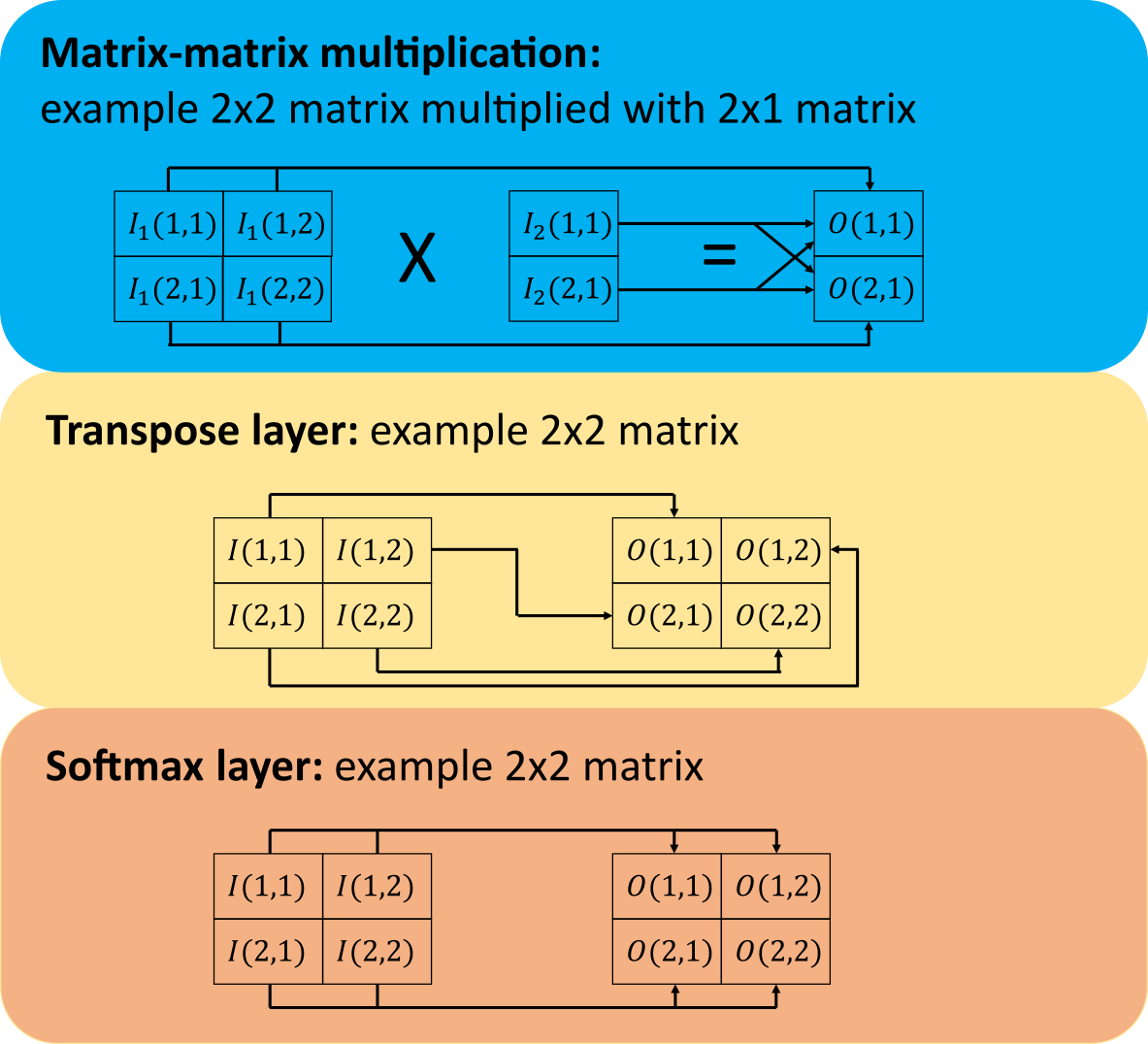}
\caption{Illustration of dependencies in different layer types.}
\label{fig:dependencies}
\vspace*{-0.3cm}
\end{figure}
In order to support layer-fused processing of transformer networks, the Stream framework is extended. To this end, support is added for both transpose and softmax layers, which are present in the attention heads of transformer networks. Specifically, this required to modify Step 1 and Step 2 of the framework.

To support the splitting of transpose and softmax layers into smaller individually-schedulable computation nodes in Step 1, we create computation nodes based on the top 'for loop' of the temporal mapping: one for each $R$ if the top 'for loop' is 'for R' etc.    

Also the computation node dependency generation of step 2 is modified to accurately model the dependencies between the computation nodes of the newly-introduced layer types in the attention heads, as illustrated in Fig. \ref{fig:dependencies}: For matrix-matrix multiplication, output position $(i,j)$ depends on the $i$-th row of the left input matrix and the $j$-th column of the right input matrix. 
For the transpose and softmax layers, additional dependency modeling is required. For a transpose layer, for each output location $(i,j)$ a dependency has to be registered towards input $(j,i)$. For a softmax layer, each output location $(i,j)$ is dependent on all input positions of row $i$, due to the summation in the denominator of the fraction. The exponent, however, is an elementwise operation, which does not need an additional dependency. The Stream framework is as such extended to automatically extract the dependency implications these additional layer types have on the individually-schedulable compute graph. This enables reuse of Stream's extensive scheduling and multicore performance modeling for transformer scheduling and mapping optimizations. No further modifications are needed for step 3, 4 and 5 as these make use of (a graph of) computation nodes, not taking into account the layer type.

\section{Validation} \label{validation}
We validate the functional correctness and accuracy of the enhanced modeling framework with the mapping of a CCT-like transformer network~\cite{hu2018cct} on the GAP8 hardware platform, a commercial Parallel Ultra Low Power (PULP~\cite{flamand2018gap})-based microcontroller developed by GreenWaves Technologies. 
The CCT model aims at detecting seizures in epileptic patients by analyzing electroencephalography, encapsulating the Integrized SoftMax, Layer Normalization and GELU from I-BERT~\cite{kim2021bert} as well as the library of optimized Attention kernel from~\cite{MCUisallyouneed}. For validation purposes, the model is deployed with a sequence length of 81 and of 128, with for each configuration 32 embedding channels and a projection space of size 32. The network is quantized to 8 bit fixed-point precision, with a linear re-quantization after every layer consisting of an element-wise multiplication, addition, and shift.
The memory hierarchy of GAP8 has 4 levels, noted L3 to L0, and each level can hold every operand type. GAP8 uses a DMA Engine to pipeline data transfers between L1 and L2 through the AXI crossbar, drastically reducing the latency per element for large transfers. Each of the 8 cores makes use of 1 MAC.
While the L2 to L1 interface is 64-bit wide, configuration and package size overhead virtually reduce the effective bandwidth to 51 bits per cycles.

\begin{figure}[btp]
\centering
\includegraphics[width=0.65\linewidth]{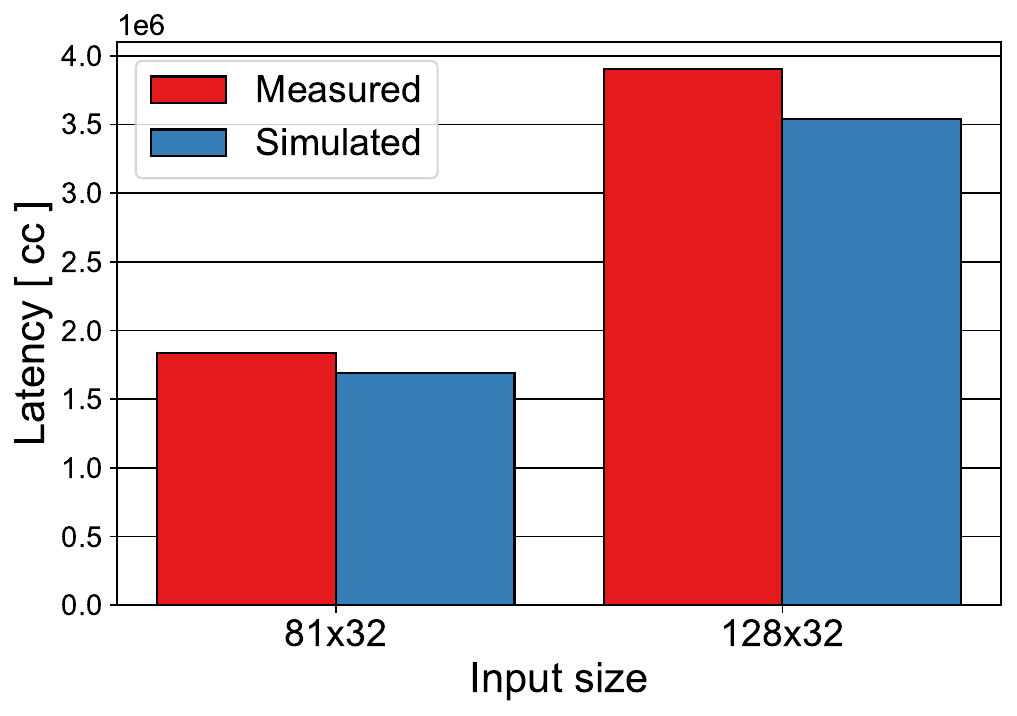}
\caption{Validation results of the framework with two different sized CCT-like networks on GAP8.}
\label{fig:validation2}
\vspace*{-0.3cm}
\end{figure}

Executing the whole MHSA on a GAP8 hardware prototype @ 100MHz is measured to take 1.836 MCycles (seq. length 81), resp. 3.905 MCycles (seq. length 128), reaching an average of 3.2 MAC/cycle. To validate the modified Stream framework, we model the described hardware architecture in Stream, as well as the mapped network execution schedules. Stream suggests a layer-fused execution, just like the used scheduling in the measurements.  The modeling framework estimates a latency of 1.692 MCycles (seq. length 81), resp. 3.540 MCycles (seq. length 128), as shown in Fig.~\ref{fig:validation2}, which is a deviation of 8\%, resp. 9\% in comparison with the real-world measurements, which is small enough to distinguish relevant cost differences between various scheduling and hardware architecture options. 

\section{Exploration with analysis} \label{exploration}
\begin{figure}
  \centering
  \begin{tabular}{@{}c@{}}
    \includegraphics[width=1.0\linewidth]{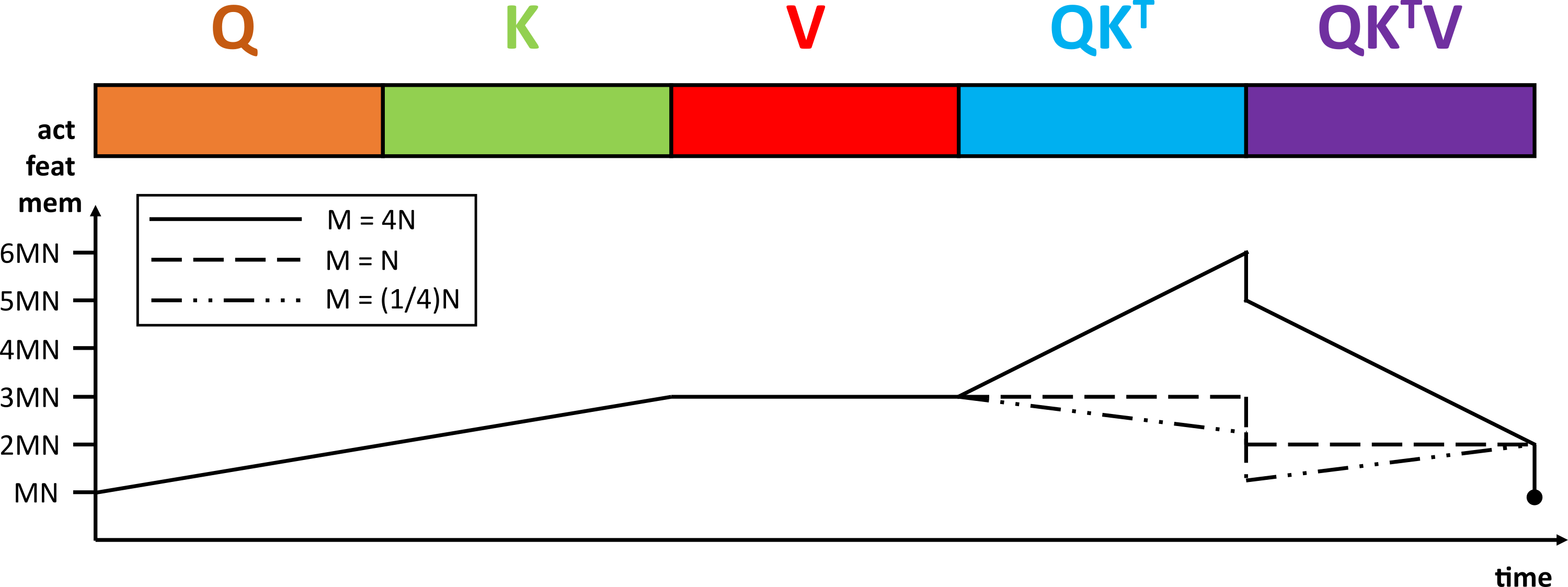} \\[\abovecaptionskip]
    \small (a) Optimized scheduling for layer-by-layer
  \end{tabular}

  \vspace{\floatsep}

  \begin{tabular}{@{}c@{}}
    \includegraphics[width=1.0\linewidth]{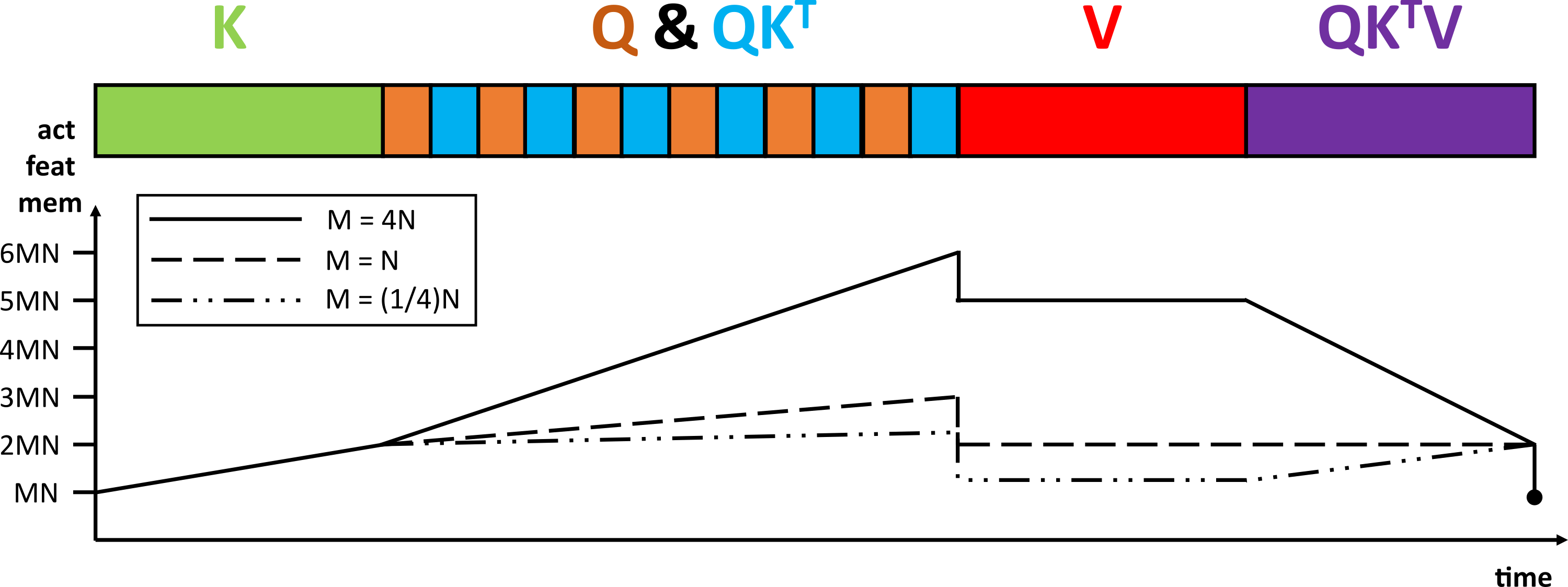} \\[\abovecaptionskip]
    \small (b) Optimized scheduling for layer-fused, with $M < N$
  \end{tabular}

  \vspace{\floatsep}

  \begin{tabular}{@{}c@{}}
    \includegraphics[width=1.0\linewidth]{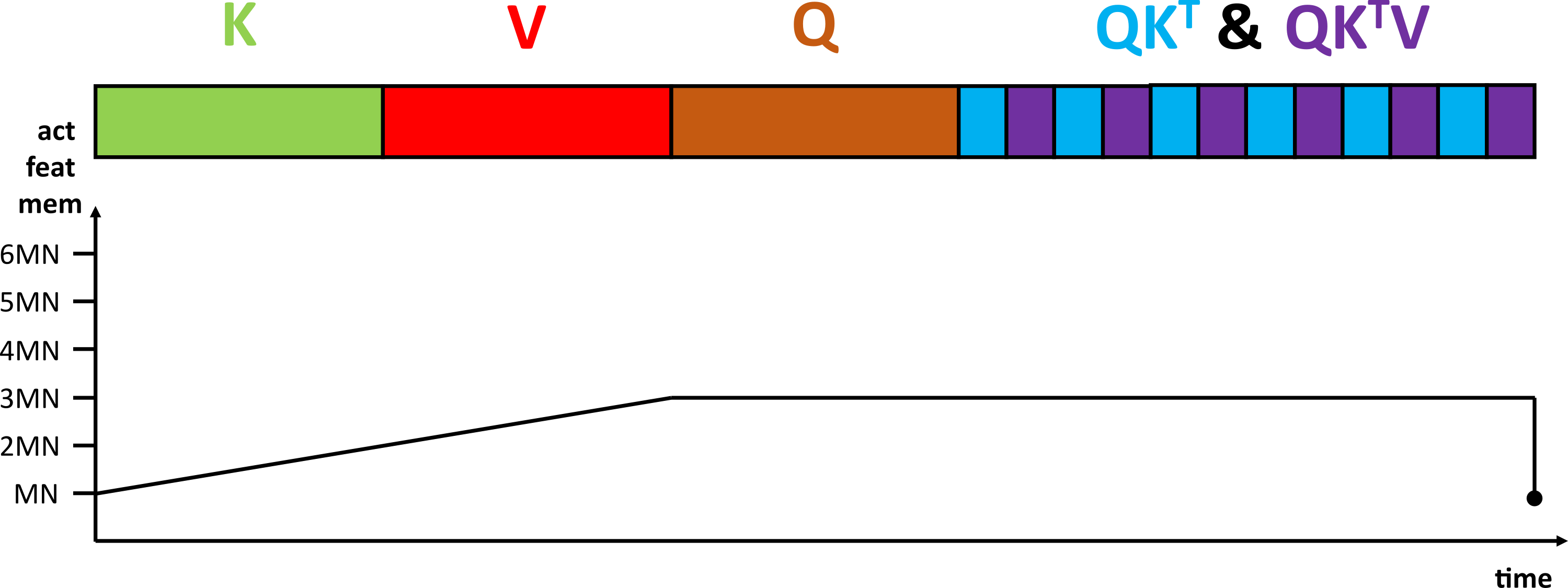} \\[\abovecaptionskip]
    \small (c) Optimized scheduling for layer-fused, with $M > N$
  \end{tabular}

  \caption{Scheduling and total active features memory usage across time for memory-optimized layer-by-layer and layer-fused execution, with best possible latency. For the layer-fused execution, we make a distinction between $M < N$ and $M > N$ as this leads to different optimal schedulings. For each scheduling, both $M > N$ as $M = N$ as $M < N$ are shown to illustrate why an optimal scheduling for one input size is not necessarily the best one for another scheduling. The dot at the end of the plots indicates that the output of the attention head should remain active to consume as the input of the next attention head. For a), the computations of $V$ and $Q.K^T$ can be swapped, this doesn't change latency or maximal memory requirement. For c), it is also an option to fuse $Q$, $Q.K^T$ and $Q.K^T.V$, instead of doing first $Q$ completely and then only fuse $Q.K^T$ and $Q.K^T.V$.} \label{summaryschedulings}
\end{figure}
\subsection{Introduction}
For typical neural networks, layer-fused execution has been shown to outperform layer-by-layer execution in terms of latency and active feature memory \cite{symons2022towards} \cite{goetschalckx2019breaking}.
In this exploration section, we will use Stream to investigate whether these benefits are still true for attention heads in transformer networks. Therefore, we will first explore the optimal layer-by-layer schedule towards the lowest latency, resp. lowest memory footprint. 
Next, results are compared to an optimized schedule using layer-fused scheduling. The study will first be conducted on a single core hardware architecture with a 64x64 array of processing elements, after which multi-core hardware mappings are explored. Memory hierarchy will be optimized to have optimal latency and memory footprint for the layer-by-layer schedule. The scheduler is given full flexibility in terms of allowed spatial and temporal unrolling.

To explore the impact of the attention head's input matrix size, we will perform all experiments with three possible input tensor shapes of equal total size (same $M.N$): square ($M = N$), deep rectangle ($M > N$, here $M=4N$) and flat rectangle ($M < N$, here $M=(1/4)N$). $M$ and $N$ are here always assumed to be multiples of the PE array size (64). The conclusions described below are exactly the same for each set of $M$ and $N$ meeting these requirements.

\subsection{Layer-by-layer Optimized Scheduling}
\subsubsection{Hardware and Mapping Exploration} 
The optimal schedule spatially unrolls (parallel hardware mapping) the $S$ and $T$ dimensions, while it moves the R dimension to the outer temporal unrolling dimension.
In this way, one row of the left input matrix can be discarded and substituted by one row of the output matrix reducing the algorithm's memory footprint. 
Therefore, we make use of two L1 memories. One for the left input matrix and output matrix (bandwidth of 64 words sufficient based on scheduling), and one for the right input matrix with a multi-banked bandwidth of 4096 words.
Each PE also has a small register file for left inputs, right inputs and outputs, respectively. Connections between these register files are also placed to make it possible to consume outputs of a given attention head layer immediately as input of a next attention head layer.
A small SIMD core is placed in parallel with the 64x64 core to compute the output of the softmax layer.

\subsubsection{Memory Footprint Analysis}
Fig.~\ref{summaryschedulings}a visualizes the memory footprint of the memory-optimal scheduling 
result obtained by the extended version of Stream for various input shapes. Probing deeper, following analysis can be made.
At the beginning of the schedule, only the input matrix is active data. The memory requirement is therefore $M.N$ words. During the computation of the $Q$ and $K$ matrix, new active features are generated without being able to discard features as we still need the input matrix to compute the $V$ matrix. Therefore, the total active memory requirement grows to $3M.N$. Next, during the computation of $V$, each iteration of the 'for R' loop results in a new row of $V$, such that a row of the input matrix with the same size can be discarded. Therefore, the active feature memory remains constant during the execution of this layer.

After consuming one row of the $Q$ matrix, this row can be discarded and substituted by a row of $Q.K^T$. Note that the $K$ or $K^T$ matrix has to be stored completely until the last input row of $Q$ is consumed. The active feature memory requirement at the end of the computation of the $Q.K^T$ matrix is $2M.N + M^2$, which is a decrease in memory if $M < N$, a constant if $M = N$ and an increase if $M > N$. After the computation of $Q.K^T$, the $K$ matrix can be discarded. Afterwards, each row of the $Q.K^T$ matrix is substituted by a row of the output matrix. When the output matrix of the attention head is computed, $V$ can be thrown away and only the $M.N$ words from the output matrix remain to be used as input of another attention head.

We can hence conclude that in the optimal layer-by-layer schedule, the maximal total active features memory requirement $A_{LBL}$ equals:
\begin{itemize}
\item If $M <= N$: $A_{LBL} = 3M.N$
\item If $M > N$: $A_{LBL} = 2M.N + M^2$
\end{itemize}

\subsection{Layer-fused Optimized Scheduling}
In this section, we use the extended version of Stream to find a layer-fused scheduling with the same optimal latency and a lower total active features memory footprint compared to the memory-optimal layer-by-layer schedule.

\subsubsection{Optimization for $M < N$} 
Fig. \ref{summaryschedulings}b visualizes the optimal execution schedule for the $M < N$ scenario, together with the corresponding memory footprint.
As can be observed, the optimized schedule fuses the computation of $Q$ and $Q.K^T$. This means that the outputs of $Q$ are immediately consumed by the computation of $Q.K^T$ and are not stored in L1 memory.

The start of the memory usage is the same as in the layer-by-layer case with an increase to $2M.N$ active feature words. Afterwards, the fusion of $Q$ and $Q.K^T$ takes place which means that the total active features memory requirement increases with the shape of $Q.K^T$ (and not $Q$) and becomes therefore equal to $2M.N + M^2$, whereafter the $K^T$ matrix can be discarded. Memory requirements during the next two layer executions can be derived using the same methodology, but do not become larger than $2M.N + M^2$. Therefore, the maximal total active features memory requirement for layer-fused execution $A_{LF} = 2M.N + M^2$.

We can now compare $A_{LBL}$ with $A_{LF}$ and define the relative memory footprint gain $\alpha = \frac{A_{LF}}{A_{LBL}}$.
For $M < N$,
\begin{equation}
\alpha = \frac{2M.N + M^2}{3M.N} = \frac{2N + M}{3N} < 1 
\end{equation}
resulting in a layer-fused scheduling with the same latency and better maximal active features memory requirement. For example, with a 128x1024 input size, $\alpha \approx 0.711$, leading to a 29\% memory requirement reduction. In the limit, we obtain
\begin{equation}
\lim_{\frac{M}{N}\to 0} \frac{2N + M}{3N}  = \frac{2}{3}
\end{equation}
which means we can reduce the total active features memory requirement with 1/3 with constant latency.\\
The same schedule, however, does not benefit execution with different input tensor sizes, as for $M > N$,
\begin{equation}
\alpha = \frac{2M.N + M^2}{2M.N + M^2} = 1
\end{equation}
and for $M = N$,
\begin{equation}
\alpha = \frac{2M.N + M^2}{3M.N} = \frac{3M^2}{3M^2} = 1
\end{equation}
Yet, other optimizations allows to execute such attention heads more efficiently, as discussed in the next paragraph. 

\subsubsection{Optimization for $M > N$}
The schedule of Fig. \ref{summaryschedulings}c is found when minimizing the feature memory requirement for attention heads with $M > N$. The memory graph now only contains one line as memory requirements are the same for all three input cases ($M > N$, $M = N$ and $M < N$). First, $K$, $V$ and $Q$ are computed in a layer-by-layer schedule. At this point, the memory requirement is $3M.N$, similar to the layer-by-layer case. 
Afterwards, the computations of $Q.K^T$ and $Q.K^T.V$ are fused, together with the softmax operation in the SIMD core. This means that features from $Q.K^T$ are not stored in local memory and one row of the $Q$ matrix is substituted by one row of the $Q.K^T.V$ matrix, with the same size. The maximal active features memory requirement, therefore, remains constant with $A_{LF} = 3M.N$ words. 

If we now again compute the relative memory footprint gain $\alpha = \frac{A_{LF}}{A_{LBL}}$ for $M > N$,
\begin{equation}
\alpha = \frac{3M.N}{2M.N + M^2} = \frac{3N}{2N + M} < 1
\end{equation}
resulting in a more memory efficient layer-fused scheduling. For example, with a 1024x128 input size, $\alpha = 0.3$, leading to a 70\% memory requirement reduction. In the limit, we obtain
\begin{equation}
\lim_{\frac{M}{N}\to\infty} \frac{3N}{2N + M}  \approx \frac{3N}{M}
\end{equation}
which means the total active features memory requirement can be reduced to a third of $\frac{M}{N}$. 
For $M <= N $,
\begin{equation}
\alpha = \frac{3M.N}{3M.N} = 1
\end{equation}
indicating that this schedule is not beneficial over layer-by-layer scheduling for those input tensor sizes.

\subsubsection{Take-away Message and State-of-the-Art Comparison}
\begin{figure}[btp]
\centering
\includegraphics[width=1\linewidth]{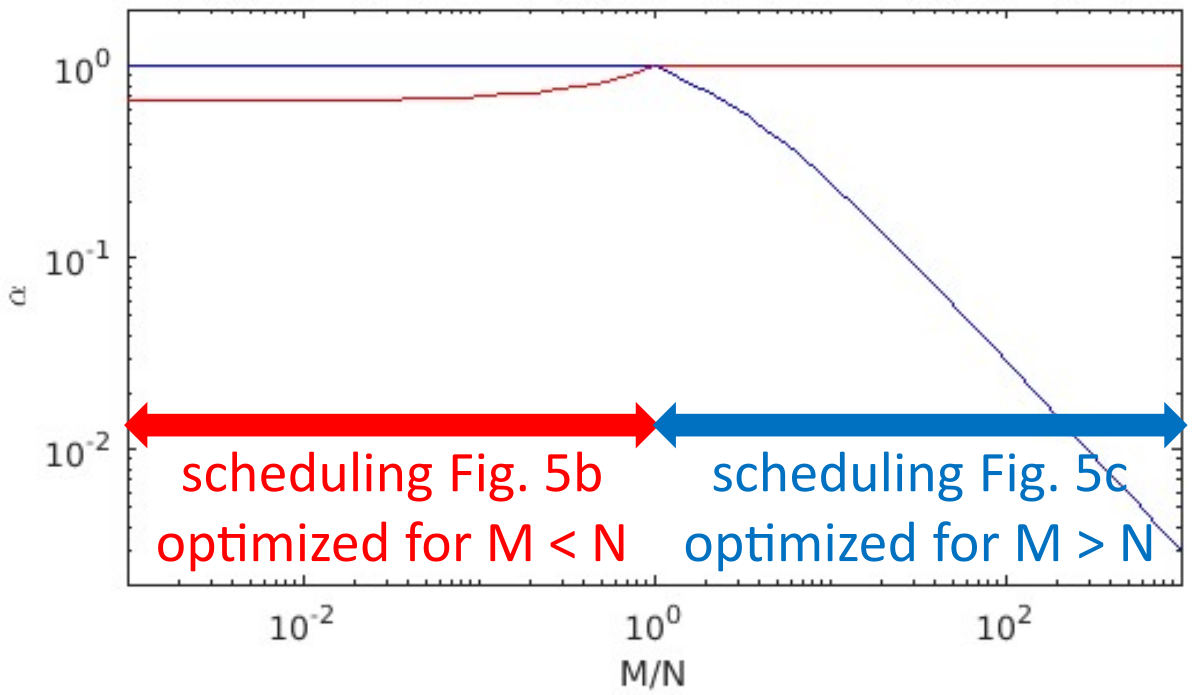}
\caption{Relative memory footprint gain $\alpha = \frac{A_{LF}}{A_{LBL}}$ in function of $\frac{M}{N}$.}
\label{fig:alpha}
\vspace*{-0.3cm}
\end{figure}

The relative memory footprint gain $\alpha$ between the layer-fused memory footprint and the layer-by-layer memory footprint is $\frac{2N + M}{3N}$ for $M < N$ and $\frac{3N}{2N + M}$ for $M > N$. The value of $\alpha$ in function of $\frac{M}{N}$ is plotted in Fig. \ref{fig:alpha}. Important to note also is that a reduction in required feature memory size will also result in a potential reduction of the total energy consumption, as a smaller memory reduces the read/write energy cost.

The optimal execution schedule differs for both cases, with a different set of layers that have to be fused. In both cases, the data of the largest matrix were immediately consumed as inputs of the next layer, with smaller output dimensions. For $M < N$, the $Q.K^T$ matrix is smaller than the $Q$ matrix. For $M > N$, the $Q.K^T.V$ matrix is smaller than the $Q.K^T$ matrix. This is also the reason why layer-fused execution doesn't reduce the memory requirements for the case where $M = N$ as all the output matrices of the 5 matrix-matrix multiplication layers have the same size. For multi-headed attention blocks on multi-core architectures, the gain is exactly the same as the one in Fig. \ref{fig:alpha} because each core executes another attention head in parallel, as no inputs or weights are typically shared among heads. Table \ref{table:sota} compares this work to previous state-of-the-art modelling frameworks, where our extensions of \cite{symons2022towards} enable transformer exploration using layer fusion on multi-accelerator architectures.

\begin{table}[h!]
\centering
\begin{tabular}{ c | c | c | c | c | c | c} 
  &  \cite{parashar2019timeloop} &  \cite{mei2021zigzag} & \cite{mei2022defines} & \cite{kwon2021heterogeneous} & \cite{symons2022towards} & \textbf{This work} \\ \hline
 Layer fusion &  $\times$ &  $\times$ & \checkmark &  $\times$ & \checkmark & \textbf{\checkmark} \\  
 Multi-accelerator &  $\times$ &  $\times$  &  $\times$ & \checkmark & \checkmark & \textbf{\checkmark}\\
 Transformer support &  $\times$ &  $\times$ &  $\times$ &  $\times$ &  $\times$ & \textbf{\checkmark} \\ [1ex] 
 \hline
\end{tabular}
\caption{Comparison with state-of-the-art modelling frameworks.}
\label{table:sota}
\end{table}

\section{Conclusion}
Transformer networks contain new layer types such as matrix-matrix multiplications with two feature matrices, transpose layers and softmax layers leading to new hardware acceleration challenges and scheduling opportunities. To explore these methodologically, this work extends Stream, a state-of-the-art multi-accelerator mapping optimization framework, with these layer types and new execution schedules. After validation, a scheduling exploration is performed from which we conclude that, depending on the size of the input matrix of the transformer's attention head, layer fusion can outperform layer-by-layer scheduling with a drastically reduced feature memory footprint, without latency penalty. For flat and wide input matrices ($M << N$), memory footprint can be reduced up to a third of the original size, while for deep and shallow input matrices ($M >> N$), benefits can get to a third of $\frac{M}{N}$. This memory reduction technique can aditionally lead to an energy reduction as smaller memories reduce the memory access energy. Both input dimension cases are characterized by a different optimal layer-fused schedule, found by the extended Stream framework. The framework, including extensions, is open-sourced at \href{https://www.github.com/kuleuven-micas/stream}{github.com/kuleuven-micas/stream}.


\section{Acknowledgment}
This project has been partly funded by the European Research Council (ERC) under grant agreement No. 101088865, the European Commission through the project CONVOLVE (101070374), the Flanders AI Research Program and KU Leuven.

\bibliographystyle{IEEEtran}
\bibliography{refs}

\end{document}